\documentclass{webofc}
\usepackage[varg]{txfonts}   
\usepackage[mathscr,scaled=1.15]{urwchancal}
\DeclareFontFamily{OT1}{pzc}{}
\DeclareFontShape{OT1}{pzc}{m}{it}%
{<-> s * [1.15] pzcmi7t}{}
\DeclareMathAlphabet{\mathpzc}{OT1}{pzc}{m}{it}

\woctitle{$13^{\rm th}$ International Conference on Meson-Nucleon Physics and the Structure of the Nucleon (MENU 2013)}
\begin{document}
\title{Images of the Origin of Mass}
%
%

\author{Craig D. Roberts\inst{1}\fnsep\thanks{\email{cdroberts@anl.gov}}
}

\institute{Physics Division, Argonne National Laboratory, Argonne, Illinois 60439, USA}

\abstract{%
Recent years have brought considerable progress with studies of the bound-state problem in continuum QCD.  A small part of that made with Dyson Schwinger equations is highlighted herein.  Topics covered include: opportunities provided by precision experimental studies of the (far) valence region; and capitalising upon new data on hadron elastic and transition form factors.
}
\maketitle
\section{Introduction}
\label{intro}
The international hadron physics programme for the coming decade can be viewed within the context of three overarching challenges: discover the meaning of confinement; determine its connection with dynamical chiral symmetry breaking (DCSB); and elucidate the signals of these phenomena in observables, so that experiment and theory together can map the nonperturbative behaviour of the strongly interacting piece of the Standard Model.  
Many paths are being followed in order to address these grand challenges but herein I focus on just two.  The community is exploiting opportunities provided by precision experimental studies of the (far) valence region, and producing theoretical computations of distribution-functions and -amplitudes with a traceable connection to QCD \cite{Holt:2010vj,Holt:2012gg}.  Such computation is critical because without it, no amount of data can reveal anything about the theory underlying the phenomena of strong interaction physics.  It is also capitalising upon new data on hadron elastic and transition form factors \cite{Holt:2012gg,Arrington:2006zm,Perdrisat:2006hj,Aznauryan:2012baS}, in order to, e.g.: chart the infrared evolution of QCD's running coupling and dressed-masses; reveal correlations that are key to baryon structure; and expose the facts and fallacies in modern descriptions of hadron structure.

The theory in which we're interested is QCD, for which there is no confirmed breakdown over an enormous energy domain: $0 < E < 8\,$TeV.  Consequently, QCD is plausibly the only known instance of a quantum field theory that can rigorously be defined nonperturbatively, in which case it is truly a field theory \emph{not} merely an effective field theory.  The possibility that QCD might be rigorously well defined is one of its deepest fascinations.  In that case, QCD might stand alone as an archetype -- the only internally consistent quantum field theory which is defined at all energy scales.  This is a remarkable possibility with wide-ranging consequences and opportunities; e.g., it means that QCD-like theories provide a viable paradigm for extending the Standard Model to greater scales than those already probed.  Contemporary research in this direction is typified by the notion of extended technicolour \cite{Sannino:2013wla}, in which electroweak symmetry breaks via a fermion bilinear operator in a strongly-interacting non-Abelian theory; and the Higgs sector of the Standard Model becomes an effective description of a more fundamental fermionic theory, similar to the Ginzburg-Landau theory of superconductivity.

Returning to the question of confinement, it is actually crucial to \emph{define} the subject; a problem canvassed in Sec.\,2.2 of Ref.\,\cite{Cloet:2013jya}, which explains that the potential between infinitely-heavy quarks measured in numerical simulations of quenched lattice-regularised QCD -- the so-called static potential -- is simply \emph{irrelevant} to the question of confinement in a universe in which light quarks are ubiquitous.  It is a basic feature of QCD that light-particle creation and annihilation effects are essentially nonperturbative and therefore it is impossible in principle to compute a quantum mechanical potential between two light quarks \cite{Chang:2009aeS}.  This means there is no flux tube in a universe with light quarks and consequently that the flux tube is not the correct paradigm for confinement.  An alternative associates confinement with dramatic, dynamically-driven changes in the analytic structure of QCD's propagators and vertices (QCD's Schwinger functions).  This translation of the confinement problem brings it into a domain that can be addressed via a concerted effort in experiment and theory: theory can identify signatures for such effects in observables and experiment can test the predictions.

Whilst the essence of confinement is still being debated, DCSB; namely, the generation of mass \emph{from nothing}, is a theoretically-established nonperturbative feature of QCD.  It is worth insisting on the term ``dynamical'' as distinct from spontaneous because nothing is added to QCD in order to effect this remarkable outcome.  Instead, simply through quantising the classical chromodynamics of massless gluons and quarks, a large mass-scale is generated.  DCSB is the most important mass generating mechanism for visible matter in the Universe, being responsible for $\approx 98$\% of the proton's mass.

The most fundamental expression of DCSB is the behaviour of the dressed-quark mass-function, $M(p)$,\footnote{Some contemporary textbooks and practitioners continue to conflate DCSB with the existence of a spacetime-independent $\bar q q$ condensate that permeates the Universe.  Whilst this might sometimes be convenient, the reality is very different.  Indeed, if quark-hadron duality is a fact in QCD, then such condensates are actually contained wholly within hadrons; i.e., they are a property of hadrons themselves and expressed, e.g., in their Bethe-Salpeter or light-front wave functions \protect\cite{Brodsky:2012ku}.}
which is a basic element in the dressed-quark propagator
\begin{equation}
\label{SgeneralN}
S(p) = -i\gamma\cdot p \, \sigma_V(p^2) + \sigma_S(p^2) =
1/[i\gamma\cdot p A(p^2) + B(p^2)] = Z(p^2)/[i\gamma\cdot p + M(p^2)]
\end{equation}
that may be obtained as a solution to QCD's fermion gap equation.  The highly nontrivial behaviour of the mass function (illustrated, e.g., in Fig.\,2.3 of Ref.\,\cite{Cloet:2013jya}) arises primarily because a dense cloud of gluons comes to clothe a low-momentum quark; and explains how an almost-massless, parton-like quark at high energies transforms, at low energies, into a constituent-like quark that possesses an effective ``spectrum mass'' $M_D \sim m_p/3$.  Consequently, the proton's mass would remain almost unchanged even if the current-quarks were truly massless.

\section{Parton structure of hadrons}
\label{sec-1}
Since the advent of the parton model and the first deep inelastic scattering experiments there has been a determined effort to deduce the parton distribution functions (PDFs) of the most stable hadrons \cite{Holt:2010vj}.  The behavior of such distributions on the far valence domain (Bjorken-$x> 0.5$) is of particular interest because this domain is definitive of hadrons; e.g., quark content on the far valence domain is how one distinguishes between a neutron and a proton.  Indeed, all Poincar\'e-invariant properties of a hadron: baryon number, charge, flavour content, total spin, etc., are determined by the PDFs which dominate on the far valence domain.  Moreover, via QCD evolution, PDFs on the valence-quark domain determine backgrounds at the large hadron collider.  There are also other questions; e.g., regarding flavour content of an hadron's sea and whether that sea possesses an intrinsic component \cite{Brodsky:1980pb,Signal:1987gz}.  The answers to all these questions are essentially nonperturbative properties of QCD.

Recognising the significance of the far valence domain, a new generation of experiments, focused on $x\gtrsim 0.5$, is planned at JLab, and under examination in connection with Drell-Yan studies at Fermilab and a possible EIC.  Consideration is also being given to experiments aimed at measuring parton distribution functions in mesons at J-PARC.  Furthermore, at FAIR it would be possible to directly measure the Drell-Yan process from high $x$ antiquarks in the antiproton annihilating with quarks in the proton.  A spin physics program at the Nuclotron based Ion Collider fAcility (NICA), under development in Dubna, might also contribute in this effort to map PDFs on the far-valence domain.

A concentration on such measurements requires theory to move beyond merely parametrising PDFs (and distribution amplitudes (PDAs), too \cite{Chang:2013pqS,Cloet:2013ttaS,Chang:2013niaS}).  Computation within QCD-connected frameworks becomes critical because without it, no amount of data will reveal anything about the theory underlying strong interaction phenomena.  This is made clear by the example of the pion's valence-quark PDF, $u_v^\pi(x)$, in connection with which a failure of QCD was suggested following a leading-order analysis of $\pi N$ Drell-Yan measurements \cite{Conway:1989fs}.  As explained in Ref.\,\cite{Holt:2010vj}, this confusion was fostered by the application of a diverse range of models.  On the other hand, a series of QCD-connected calculations  \cite{Hecht:2000xa,Aicher:2010cb,Nguyen:2011jy} subsequently established that the leading-order analysis was misleading, so that $u_v^\pi(x)$ may now be seen as a success for the unification of nonperturbative and perturbative studies in QCD.

A framework that provides access to the pion's valence-quark PDF can also be employed to compute its PDAs.  For example, the pion's leading-twist two-particle PDA is given by the following projection of the pion's Bethe-Salpeter wave function onto the light-front \cite{Chang:2013pqS}
\begin{equation}
f_\pi\, \varphi_\pi(x) = {\rm tr}_{\rm CD}
Z_2 \! \int_{dq}^\Lambda \!\!
\delta(n\cdot q_\eta - x \,n\cdot P) \,\gamma_5\gamma\cdot n\, \chi_\pi(q;P)\,,
\label{pionPDA}
\end{equation}
where: $f_\pi$ is the pion's leptonic decay constant; $\int_{dq}^\Lambda$ is a Poincar\'e-invariant regularization of the four-dimensional integral, with $\Lambda$ the ultraviolet regularization mass-scale; $Z_{2}(\zeta,\Lambda)$ is the quark wave-function renormalisation constant, with $\zeta$ the renormalisation scale; $n$ is a light-like four-vector, $n^2=0$; $P$ is the pion's four-momentum, $P^2=-m_\pi^2$ and $n\cdot P = -m_\pi$, with $m_\pi$ being the pion's mass; and $\chi_\pi(q;P)$ is the pion's Poincar\'e covariant Bethe-Salpeter wave function.

\begin{figure}[t]
\centering
\sidecaption
\includegraphics[width=0.5\textwidth,clip]{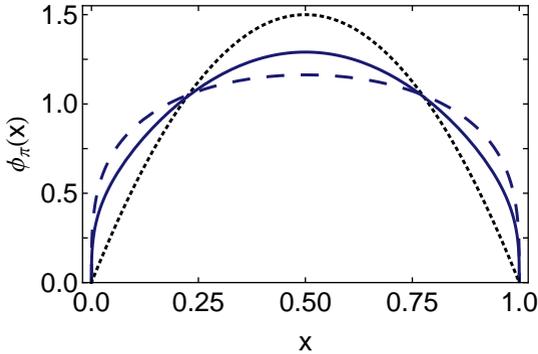}
\caption{\label{FigpionPDA}
Twist-two pion PDA computed using two vastly different DSE truncations at a scale $\zeta=2\,$GeV. \emph{Dashed curve} -- rainbow-ladder (RL), the leading-order in a systematic, sym\-metry-preserving scheme \protect\cite{Munczek:1994zz,Bender:1996bb}; and \emph{solid curve} -- the most sophisticated kernel that is currently available; namely, a DCSB-improved (DB) kernel that incorporates essentially nonperturbative effects generated by DCSB that are omitted in RL truncation and any stepwise improvement thereof \cite{Chang:2009zb,Chang:2010hb,Chang:2011ei}.  These results are consistent with contemporary lattice-QCD \cite{Cloet:2013ttaS,Segovia:2013ecaS}.  The dotted curve is $\varphi^{\rm asy}_\pi(x)$, the result obtained in conformal QCD \cite{Efremov:1979qk,Lepage:1979zb}.}
\end{figure}

The amplitude in Eq.\,\eqref{pionPDA} has been computed using two very different truncations of QCD's DSEs \cite{Chang:2013pqS}, with the result depicted in Fig.\,\ref{FigpionPDA}.  Both kernels agree: compared with the asymptotic form, there is a marked broadening of $\varphi_\pi(x)$, which owes exclusively to DCSB.  This causal connection may be claimed because the PDA is computed at a low renormalisation scale in the chiral limit, whereat the quark mass function owes entirely to DCSB.  Hence, in Fig.\,\ref{FigpionPDA} one has, for the first time, exposed DCSB on the light-front.  Moreover, the dilation measures the rate at which a dressed-quark approaches the asymptotic bare-parton limit.  It can be verified empirically at JLab12; e.g., in measurements of the pion's electromagnetic form factor and the ratio of the proton's electric and magnetic form factors.

A question of more than thirty-years standing can be answered using Fig.\,\ref{FigpionPDA}; namely, when does $\varphi^{\rm asy}_\pi(x)$ provide a good approximation to the pion PDA?  Plainly, not at $\zeta=2\,$GeV.  The ERBL evolution equation \cite{Efremov:1979qk,Lepage:1979zb} describes the $\zeta$-evolution of $\varphi_\pi(x)$; and applied to $\varphi_\pi(x)$ in Fig.\,\ref{FigpionPDA}, one finds \cite{Cloet:2013ttaS,Segovia:2013ecaS} that $\varphi^{\rm asy}_\pi(x)$ is a poor approximation to the true result even at $\zeta=200\,$GeV.  Given that the pion's PDA is difficult to measure directly, it is better to work with the pion's valence-quark PDF, $u_{\rm v}^\pi(x;\zeta)$, which is known empirically \cite{Hecht:2000xa,Aicher:2010cb}.  Since the discovery of asymptotic freedom it has been known that $\varphi^{\rm asy}_\pi(x)$ can only be a good approximation to the pion's PDA when it is accurate to write $u_{\rm v}^\pi(x;\zeta) \approx \delta(x)$.  Plainly, in this case:
\begin{equation}
\lim_{\Lambda_{\rm QCD}/\zeta \to 0} \langle x \rangle^\pi_\zeta  = 0\,,\quad
\langle x \rangle^\pi_\zeta :=
\int_0^1 dx\, x \, u_{\rm v}^\pi(x;\zeta) \,;
\label{averagex}
\end{equation}
i.e., the light-front fraction of the pion's momentum carried by dressed valence-quarks is zero.

How does this compare with empirical reality?  At $\zeta=2\,$GeV, dressed valence-quarks carry 45\% of the pion's light-front momentum; i.e.  \cite{Hecht:2000xa,Aicher:2010cb}, $2\,\langle x \rangle^\pi_{\zeta_2} = 0.45$.  The remainder is carried by glue (44\%) and sea-quarks (11\%).  The equation describing the $\zeta$-evolution of $u_{\rm v}^\pi(x)$ is known.  Hence one can readily compute $\langle x \rangle^\pi_\zeta$ and find \cite{Cloet:2013jya} the momentum fraction evolves so slowly that even at LHC energies 25\% of the pion's momentum is carried by dressed-valence quarks, 21\% is carried by sea-quarks and 54\% is carried by glue.  (Owing to saturation of the gluon fraction, evolution slowly shifts momentum from the valence- to the sea-quarks.)  Plainly, even at LHC energy scales, nonperturbative effects such as DCSB are playing a crucial role in setting the scales in PDAs and PDFs.

\section{Electromagnetic structure of hadrons}
\label{sec-2}
The pion electromagnetic form factor has long enthralled hadron physicists; and in Ref.\,\cite{Maris:2000sk} a one-parameter model for QCD's RL interaction kernel was used to compute $F_\pi(Q^2)$, producing a prediction that was confirmed by subsequent JLab experiments \cite{Volmer:2000ek,Huber:2008id}.  A weakness of the study, however, was that it used brute numerical methods in the computation and was therefore limited to $Q^2\leq 4\,$GeV$^2$.  That has now changed.  Using a refinement of known methods \cite{Nakanishi:1969ph}, Ref.\,\cite{Chang:2013niaS} provided a prediction of the pion's electromagnetic form factor to arbitrarily large-$Q^2$; and, moreover, correlated that prediction with the pion PDA depicted in Fig.\,\ref{FigpionPDA} via the formula of perturbative QCD (pQCD) \cite{Efremov:1979qk,Lepage:1979zb}: with $f_\pi$ being the pion decay constant and $\alpha_s(Q^2)$, QCD's running coupling,
\begin{equation}
\label{pionUV}
\exists Q_0>\Lambda_{\rm QCD} \; |\;  Q^2 F_\pi(Q^2) \stackrel{Q^2 > Q_0^2}{\approx} 16 \pi \alpha_s(Q^2)  f_\pi^2 \mathpzc{w}_\varphi^2,
\quad \mathpzc{w}_\varphi = \frac{1}{3} \int_0^1 dx\, \frac{1}{x} \varphi_\pi(x)\,.
\end{equation}

\begin{figure}[t]
\centering
\sidecaption
\includegraphics[width=0.5\textwidth,clip]{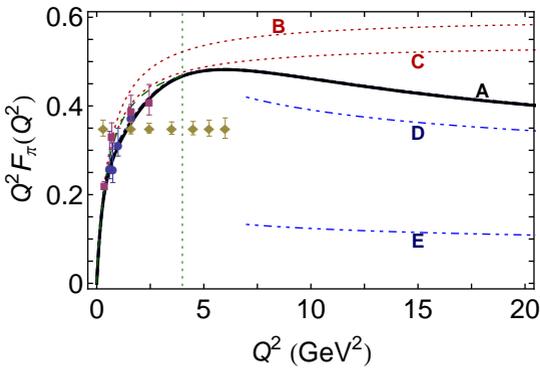}
\caption{\label{Fig2} $Q^2 F_\pi(Q^2)$.  \emph{Solid curve}\,(A) -- prediction of Ref.\,\cite{Chang:2013niaS}; and \emph{long-dashed curve} -- result in\,Ref.\,\protect\cite{Maris:2000sk}, which is limited to $Q^2<4\,$GeV$^2$ (boundary indicated by vertical dotted line).
Other curves, top to bottom: \emph{dotted}\,(B) -- monopole ``$1/(1+Q^2/m_\rho^2)$;'' \emph{dotted}\,(C) -- monopole fitted to data in Ref.\,\protect\cite{Amendolia:1986wj}, with mass-scale $0.74\,$GeV;
\emph{Dot-dot--dashed}\,(D) -- Eq.\,\protect\eqref{pionUV} computed with RL result for $\varphi_\pi(x)$ in Fig.\,\ref{FigpionPDA}; and \emph{Dot-dot--dashed}\,(E) -- Eq.\,\protect\eqref{pionUV} computed with $\varphi_\pi^{\rm asy}(x)$ in Fig.\,\ref{FigpionPDA}.
Filled-circles and -squares: data described in Ref.\,\protect\cite{Huber:2008id}. Filled diamonds indicate the projected reach and accuracy of a forthcoming experiment\,\protect\cite{E1206101}.
}
\end{figure}

Fig.\,\ref{Fig2} shows that the pQCD prediction obtained when the pion valence-quark PDA has the form appropriate to the scale accessible in modern experiments (curve-D) is markedly different from the result obtained using the asymptotic PDA (curve-E).  Moreover, the near agreement between curve-D and the modern DSE prediction (curve-A) enables one to conclude that on $Q^2 \gtrsim 8\,$GeV$^2$ one should, for the first time, see quark-counting rules and QCD scaling violations in an hadron elastic form factor and, simultaneously, via the normalisation, a clear signal for the emergent phenomenon of DCSB.

Given the enormous impact of DCSB on meson properties, it must also be included in the analysis of baryon properties.  That is readily achieved by employing a Poincar\'e-covariant Faddeev equation \cite{Cahill:1988dx}, a dynamical prediction of which is the appearance of nonpointlike quark$+$quark (diquark) correlations within the proton \cite{Cahill:1987qr}.  It should also be stressed that these dynamically generated correlations are not the pointlike diquarks of yesteryear.  The modern diquark correlation is nonpointlike, with the charge radius of a given diquark being typically 10\% larger than its mesonic analogue \cite{Roberts:2011wyS}.  Hence, diquarks are soft components within baryons.  Notably, empirical evidence in support of the presence of diquarks in the proton is accumulating
\cite{Close:1988br,Cloet:2005pp,Cates:2011pz,Wilson:2011aa,Cloet:2012cy,Qattan:2012zf}.

\begin{figure}[t]
\centering
\includegraphics[width=0.48\textwidth,clip]{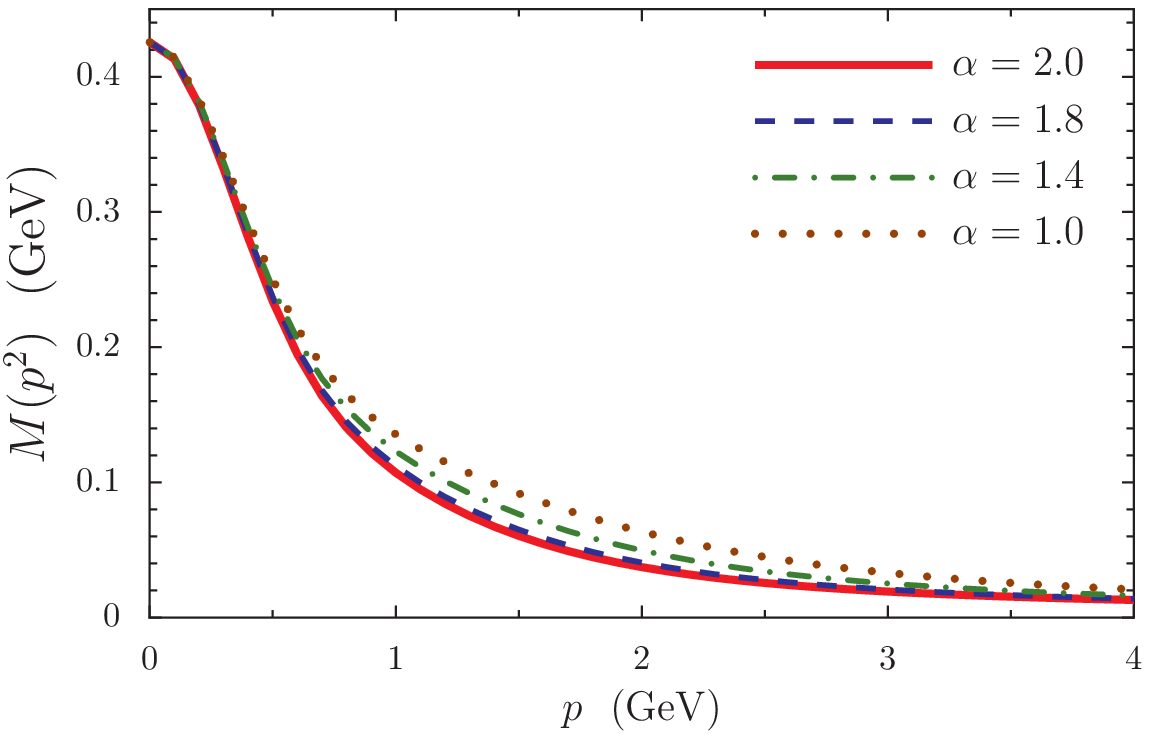}
\includegraphics[width=0.48\textwidth,clip]{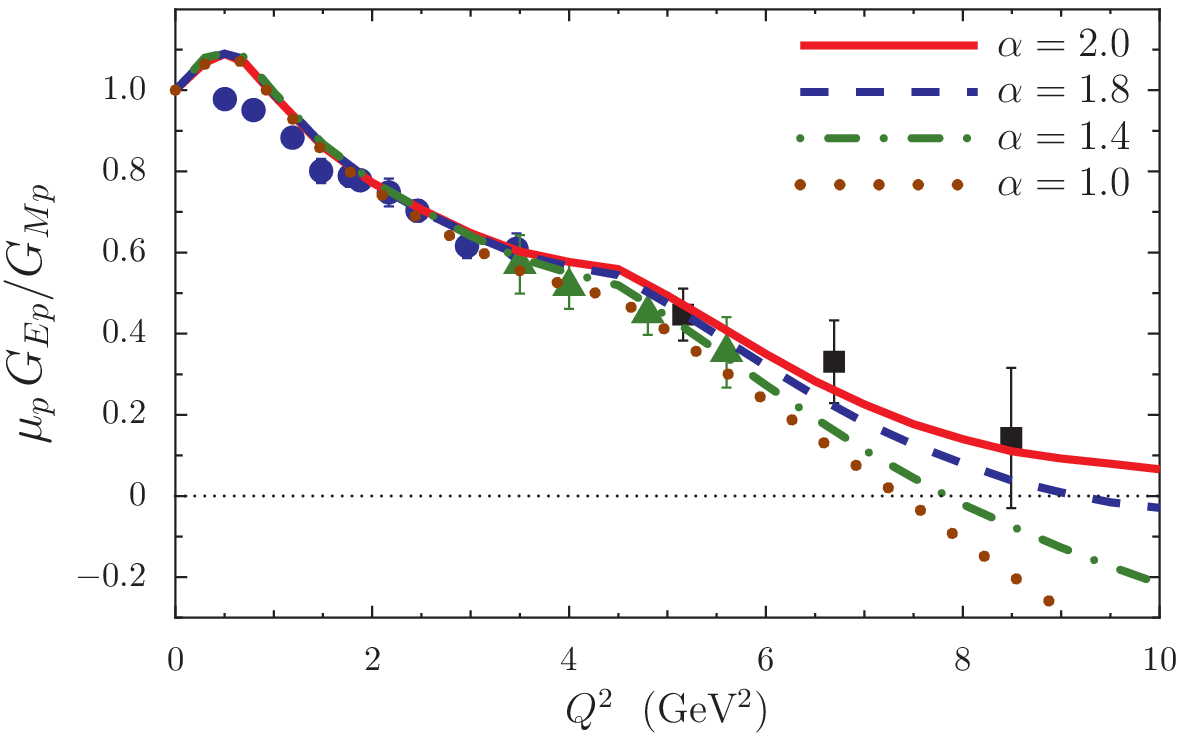}
\caption{\label{fig3}
\emph{Left panel}.  Dressed-quark mass function employed in Ref.\,\protect\cite{Cloet:2013gva}.  $\alpha=1$ specifies the reference form and increasing $\alpha$ diminishes the domain upon which DCSB is active.
\emph{Right panel}.  Response of $\mu_p G_E/G_M$ to increasing $\alpha$; i.e., to an increasingly rapid transition between constituent- and parton-like behaviour of the dressed-quarks.  Data are from Refs.\,\protect\cite{Jones:1999rz,Gayou:2001qd,Gayou:2001qtS,
Punjabi:2005wqS,Puckett:2010ac,Puckett:2011xgS}.}
\end{figure}

Recently, complementing the access that $F_\pi(Q^2)$ provides to DCSB, an interesting effect was exposed in the ratio $\mu_p G_E^p/G_M^p$ \cite{Cloet:2013gva}.  Namely, as illustrated in Fig.\,\ref{fig3}, apparently small changes in the dressed-quark mass function, $M(p)$ in Eq.\,\eqref{SgeneralN}, within the domain $1 < p({\rm GeV})<3$ have a striking effect on the proton's electric form factor.  As explained in Ref.\,\cite{Cloet:2013gva}, the possible existence and location of the zero is determined by behaviour of $Q^2F_2^p(Q^2)$, where $F_2^p$ is the proton's Pauli form factor and, like the pion's PDA, $Q^2F_2^p(Q^2)$ measures the rate at which dressed-quarks become parton-like: $F_2^p \equiv 0$ for bare quark-partons.  Therefore, $G_E$ can't be zero on the bare-parton domain.  It follows that the possible existence and location of a zero in the ratio $\mu_p G_E^p/G_M^p$ are a direct measure of the nature of the quark-quark interaction in the Standard Model.  This also leads to the prediction:
\begin{equation}
G_E^n(Q^2) > G_E^p(Q^2)  \; \mbox{on} \; Q^2 > 4{\rm GeV}^2.
\end{equation}

A great deal more information is contained in nucleon elastic form factors.  This may be illustrated by exhibiting a connection with valence-quark PDFs at very high $x$. Indeed, the endpoint of the far valence domain, $x=1$, is especially significant because, whilst all familiar PDFs vanish at $x=1$, ratios of any two need not; and, under DGLAP evolution, the value of such a ratio is invariant \cite{Holt:2010vj}.  Thus, e.g., with $d_v(x)$, $u_v(x)$ the proton's $d$, $u$ valence-quark PDFs, the value of $\lim_{x\to 1} d_v(x)/u_v(x)$ is an unambiguous, scale invariant, nonperturbative feature of QCD.  It is therefore a keen discriminator between frameworks that claim to explain nucleon structure.  Furthermore, Bjorken-$x=1$ corresponds strictly to the situation in which the invariant mass of the hadronic final state is precisely that of the target; viz., elastic scattering.  The structure functions inferred experimentally on the neighborhood $x\simeq 1$ are therefore determined theoretically by the target's elastic form factors.

This connection was used by Refs.\,\cite{Wilson:2011aa,Roberts:2013mja} to deduce formulae, expressed in terms of diquark appearance and mixing probabilities, from which one may compute ratios of unpolarised and also longitudinal-spin-dependent $u$- and $d$-quark parton distribution functions on $x\simeq 1$.  Through a comparison with predictions from other approaches plus consideration of extant and planned experiments, Ref.\,\cite{Roberts:2013mja} showed that the measurement of nucleon longitudinal spin asymmetries on $x \gtrsim 0.8$ can add enormously to our capacity for discriminating between contemporary pictures of nucleon structure.

\section{Epilogue}
The physics of hadrons is unique.  Its study brings experiment and theory into contact with QCD, a fundamental theory of astonishing complexity in which the elementary degrees-of-freedom are intangible and only composites reach detectors.  Confinement is one of QCD's distinguishing features.  However, there is not even an agreed theoretical definition.  Dynamical chiral symmetry breaking, on the other hand, is understood, and it is crucial to any understanding of hadron phenomena.  It appears highly likely that these phenomena are intimately related, and share the same origin and fate.  One thing is certain, experimental and theoretical study of the bound-state problem in continuum QCD promises to provide many more predictions, insights and answers.

\begin{acknowledgement}
The material described in this contribution is the result of an international collaborative effort that has involved, amongst others,
A.~Bashir,
S.\,J.~Brodsky,
L.~Chang,
C.~Chen,
I.\,C.~Clo{\"e}t,
J.\,J.~Cobos-Martinez,
B.~El-Bennich,
X.\,L.~Guti\'errez-Guerrero,
G.~Krein,
R.\,J.~Holt,
Y.-X.~Liu,
S.-X.~Qin,
A.~Raya,
H.\,L.\,L.~Roberts,
S.\,M.~Schmidt
J.~Segovia,
R.~Shrock,
P.\,C.~Tandy,
A.\,W.~Thomas,
S.-L.~Wan,
D.\,J.~Wilson
and
H.-S.~Zong.
This work was supported by an \emph{International Fellow Award} from the Helmholtz Association; and Department of Energy, Office of Nuclear Physics, contract no.~DE-AC02-06CH11357.
\end{acknowledgement}
%

\end{document}